\documentclass[aps,prl,preprint,showpacs,floatfix]{revtex4}
\usepackage{graphicx}

\begin{document}

\title{Role of Correlation on Charge  Carrier  Transport  in  Organic Molecular Semiconductors} 

\author{Ajit Kumar Mahapatro and Subhasis Ghosh}

\affiliation{School of Physical Sciences, Jawaharlal Nehru University,
New Delhi 110067, India}

\vspace{0.5in}

\begin{abstract}

We have investigated the charge carrier  transport in organic  molecular semiconductors. It has been found that mobility is a  function of electric field and temperature due to hopping conduction. Several theoretical models for charge transport in disordered solids have been debated over the role of spatial and energetic correlation in these systems and  such correlations have been recently shown to explain the universal electric field dependence of mobility. We have compared and evaluated the applicability of different theoretically proposed models using very simple experimental results and based on our extensive analysis, we have found that correlation is important to explain the electrical transport in these systems.
\end{abstract}


\maketitle

\newpage

Recently, considerable debate\cite{hb93,dhd96,svn98} is going on regarding the charge carrier transport in organic molecular semiconductors(OMS), which belong to wide class of material, known as disordered organic molecular solids(DOMS), because of two reasons: first, their application in display devices and second, the fundamental understanding of charge transport in these materials.  There are several distinguishing features of these solids, (i) they are composed of organic molecules, held together loosely by   weak van der Waals type intermolecular coupling while the intramolecular coupling is strong; (ii) the absence of long range order in these disordered materials lead to the localization of the electronic wave function and the formation of a broad Gaussian density of states(GDOS) and  the most important one is (iii) the carrier mobility $\mu$ exhibits a nearly universal Poole-Frenkel(PF) behavior\cite{dhd96}

\begin{equation}
\mu(F,T)=\mu(0,T) \exp\left(\gamma(T) \sqrt{F}\right)
\end{equation}

\noindent where  $\mu(0,T)$ and $\gamma(T)$ are temperature dependent quantities, known as the zero field mobility and the field activation of the mobility, respectively.  In DOMS,  hopping among the molecular sites having comparable energies, describes the transport of charge carriers through the GDOS of highest occupied molecular orbital(HOMO) and/or lowest unoccupied molecular orbital(LUMO). 
   Recently, several theoretical\cite{hb93,dhd96,svn98} models for charge transport in disordered solids have been debated over the role of spatial and energetic correlation in these systems and  such correlations have been recently shown to explain the universal electric field dependence(Eq.1) of mobility.  In this paper, we have examined the charge carrier  transport in OMS  and we show for the first time, by means of simple experiments that  correlation is important in explaining charge carrier transport in OMS.

For our study we have chosen  metal phthalocyanine(MePc) based thin films as the  organic solids because beside their potential application for {\sl plastic} based optoelectronic devices and transistor, these materials are (i) the most chemically and thermally stable compounds, (ii) suitable for  organic single layer structures with different metal electrodes without effecting the interface properties and finally (iii) reproducibility of experimental data for devices with MePc as  organic layers, which is a major problem with most other organic solids due to their degradation with time and the interaction of organic materials with the different metal electrodes, notablly with  low work function metal electrodes. MePc based organic single layer devices were prepared by sandwiching a thin  organic layer of MePc between indium tin oxide(ITO) coated glass substrate and thermally evaporated metal electrodes, using vacuum deposition technique. We report the experimental investigation on charge carrier  transport in hole only devices based on   metal/MePc/metal structures. Work functions of ITO and Cu are  4.75eV and 4.6eV, respectively,  very close to the ionization potential(4.8eV) of MePc\cite{gp98}, hence the ITO/MePc and ITO/Cu  interfaces behave an Ohmic contact. The temperature dependent current-voltage(J-V) characteristics were studied for ITO/MePc/Al structures with different MePc layers, choosing ITO as the anode. By properly choosing contacting metals(Cu and Al), current injection and transport due to holes have been investigated. Here we have chosen  copper phthallocyanine(CuPc) and  zinc phthalocyanine(ZnPc) as the MePc for these investigations. High purity, sublimated CuPc and ZnPc have been
procured from Aldrich Chemical Co. CuPc(ZnPc) was evaporated on ITO
coated glass substrates from a resistively heated tungsten boat.
 Each layer was deposited in a
vacuum chamber  at a rate of 5-10$\AA$/s. Thickness of the
resulting organic films was between 25 to 500nm. The metals for
the top electrode(Au, Cu and Al) were placed in a resistively
heated tungsten filament and evaporated  in a vacuum chamber. All
evaporations were done  at a base pressure of 2$\times10^{-6}$
Torr. We have measured the J-V characteristics by
continuously varying the bias and also by interrupting the bias
between each voltage step and in both cases, we have found same
results. The J-V characteristics for the ITO/MePc/metal
structures with different thickness of MePc layers were
reproducible at all temperatures and showed no hysteresis.

Fig.1  shows the J-V characteristics of ITO/CuPc/Al and ITO/CuPc/Cu at room temperature. The experiment consisted of two steps. First: the current due to hole injection from ITO was measured by supplying it to positive bias(forward bias) and second: the current due to hole injection from Cu and Al was measured by reversing the polarity of the bias voltage, i.e., biasing Al and Cu electrodes positively(reverse bias). It is clear from  Fig.1 that J-V characteristics in case of ITO/CuPc/Cu display almost symmetric behavior in both cases(hole injection either from ITO, or from Cu electrodes). This is because in both cases there is small energy barriers of 0.05eV in case of ITO/MePc  and 0.2eV in case of Cu/MePc interface, giving rise to space charge limited(SCL) bulk  current when either ITO or Cu is positively biased. It has been shown\cite{akm02,psd99}, for Schottky energy barrier(SEB) less than about 0.3-0.4eV, the current flow is due to SCL\cite{peb96,ajc97,pwmb97}. But, in case of ITO/MePc/Al devices, J-V characteristics display diode like asymmetric behavior. Current density increases by almost five orders of magnitude by reversing the polarity of the bias.

In several cases\cite{gp98,peb96,pwmb97}, the current transport mechanism has been explained in non-crystalline organic semiconductor by extending the space charge limited current(SCLC)
and trap charge limited current(TCLC) transport models in crystalline semiconductors\cite{ar55,pm62} and J-V characteristics beyond Ohm's law was given by 

\begin{equation}
J \propto \frac{V^{m+1}}{d^{2m+1}}
\end{equation} 

\noindent  where $m=1$ in case of SCLC and $m=T_0/T$ in case of TCLC, 
$T_0$ is the characteristic temperature of the traps and d is the thickness of the sample.  We found in all our single layer devices with different electrodes(Al, Cu) and
different MePcs(CuPc, ZnPc), J-V characteristics follow SCLC
like behavior($J\propto V^2$) at low bias in forward bias condition, i.e. when ITO electrode is positively biased. It is appealing to
attribute this behavior to SCLC, but as the bias increases the
slope $(m+1)$ in log-log plot of J vs. V curves increases gradually
from $m=1$. This deviation can not be explained\cite{pwmb97} as due to the
presence of traps, because SCLC would not apply for low bias
either. There are valid objections\cite{pwmb97,ai98,svr00} against the use of
transport theory(SCLC and TCLC) for crystalline systems  in case of
disordered systems, like OMS.  Failure of the proposed SCLC and/or TCLC for
crystalline insulators\cite{ar55,pm62} in our devices and several other
cases\cite{pwmb97,ai98,svr00} can be argued for following reasons. 
(i) As mentioned already, OMS does not have  valence and conduction bands(as in case of crystalline semiconductor) due to lack of long range ordering in these materials.
A distinguishing feature of molecular and polymeric organic
semiconductors is that they are molecular glasses, composed of
molecular entities held together by weak van der Walls interactions.
They are characterized with HOMO and LUMO bands with GDOS. The tail states in HOMO/LUMO generally acts as traps and get filled
with an application of bias due to the penetration of quasi-Fermi
level inside the GDOS.  (ii)  SCLC and/or TCLC  theories are applicable for  crystalline systems and  assume constant mobility.  In case of molecular semiconductors, mobility is being affected by the disorders  and the transport
takes place by electrons hopping through MePc molecular sites
having comparable energies within the GDOS resulting bulk limited
behavior in the current conduction.  In this case, mobility
becomes a function of electric field due to the distribution in
energy of hopping sites. Essentially, the electric field assists
carriers in overcoming the potential barrier between sites with
different energies. It has been found that the mobility $\mu$ of
carriers in thin films of  poly-phenylene vinylene(PPV)\cite{pwmb97,svr00} and
 tris, 8-hydroxyquinoline aluminum(Alq3)\cite{rgk95} over wide range of fields can be described according to universally observed PF relation(Eq.1). 
(iii) In several reports\cite{peb96,ajc97,pwmb97}, thickness dependence of current has been used to corroborate the SCLC theory. But it has been shown\cite{ai98} that
current transport based on field dependent mobility due to hopping
of charge carriers can account for   similar thickness dependence.
It is also possible to account for $J\propto V^m$ relation
without unphysical assumptions of TCLC theory.

The temperature dependence of electrical conductivity associated with the hopping motion in non-crystalline solids is explained by

\begin{equation}
\rho(T)=\rho_0 exp\left[\left(\frac{T_0}{T}\right)^\alpha \right]
\end{equation}

\noindent where $\rho_0$ is the temperature independent resistivity and $T_0$ is the characteristic temperature, $\alpha = \frac{n+1}{n+D+1}$ is known as hopping exponent,  D is the dimension, n is the exponent in  the density of states g(E) in the vicinity of the Fermi energy $E_F$, which behaves like
$g(E)\sim(|E-E_F|)^n$.  
For an energy independent density of states($n=0$),  the values $\alpha$  are 1/4 and 1/3 in 3-D and   2-D, respectively. This is the Mott-Davis variable range hopping\cite{nfm79} mechanism. Shklovskii and Efros\cite{ale75} have shown that   Coulomb interaction  leads to a gap in $g(E)$ around $E_F$ and  $n=1$ in 2-D, i.e, $g(E)\sim(|E-E_F|)$  and $n=2$ in 3-D, i.e, $g(E)\sim(|E-E_F|)^2$. Hence electron-electron interaction change the variable range hopping exponent to  $\alpha=\frac{1}{2}$ in both 2-D and 3-D. Temperature dependence  of resistivity $\rho(T)$  over wide range of temperature for two thicknesses of CuPC  is shown in Fig.2. We have tried to fit the resistivity data with different $\alpha$ values. Fig.2 shows the $ln(\rho)$ vs $1/T^{1/4}$ and also the $ln(\rho)$ vs $1/T^{1/2}$ plot. It is observed that the 100nm structure shows better fitting with the $1/T^{1/4}$ plot, whereas the 400nm structure shows better fitting with the $1/T^{1/2}$ plot. For an improvement upon this  {\sl eyeball} approach, we adopt the method used by Hill\cite{rmh76}. He introduces the quantity $\epsilon_a = d[ln(\rho)]/d(1/T)$ as the temperature dependent activation energy for variable range hopping conduction. Eq.3  can be expressed as  

\begin{equation}
ln \left(\frac{d[ln\rho(T)]}{d(1/T)} \right) = ln(\alpha T_0^\alpha) + (1-\alpha) ln T
\end{equation}

\noindent One can determine $\alpha$ from $\epsilon_a$ vs $T$ plot in log-log scale. We have plotted according to Eq.4, the straight line fit gives the value of $\alpha = 1/5$ for 100nm(Fig.3) and $\alpha = 3/5$ for 400nm(Fig.3) CuPc layers. Since the value 1/5 is closer to 1/4 and 3/5 is closer to 1/2, the straight line fit in $\left( \frac{1}{T}\right)^\alpha$ plots shows better fitting with $\alpha = \frac{1}{4}$ for 100nm and with $\alpha = \frac{1}{2}$ for 400nm structures. Hence, it seems hopping mechanism may be  in lower and higher thickness of CuPc layers.  Moreover, hopping mechanism may be altogether different from either Mott-Davis\cite{nfm79} or Efros-Shklovskii\cite{ale75}  in OMS and the pre-exponential factor($\rho_0$) may depend on temperature. Fig.4 shows the temperature dependent resistivity  at various electric fields for ITO/CuPc(100nm)/Al structure. The  linear dependence in $ln \rho(T)$ vs $1/T^{1/4}$ upto 30K is observed.  
The electric field dependence on resistivity  is studied for the $\rho_0$  at different electric fields. The straight line fit of $ln(\rho_0)$ vs $F^{1/2}$ plot, shown as inset in Fig.4,  signifies the universally observed Poole-Frenkle relation($ln (\mu)  \propto F^{1/2}$) concerning  charge carrier transport in OMS.

To fit the experimental results, we have modified SCLC transport theory by simultaneously solving the current density equation and Poission's equation with electric field dependent mobility(PF relation i.e. Eq.1)  
 
\begin{equation}
J(x) = p(x)e\mu\left [ F(x),T \right ] F(x) 
\end{equation}

\begin{equation}
\frac{\epsilon}{e}\frac{dF(x)}{dx} = p(x)
\end{equation}

\noindent where $p(x)$ is the hole density at position $x$ and
$\mu(F,T)$ is the hole mobility in MePc.  Taking the boundary conditions (i) ohmic contacts at ITO/MePc interface(at $x=0$) i.e. electric field vanishes($F=0$) at $x=0$,
and (ii) hole density is equivalent with the density of states($N_v$) in HOMO of  MePc, i.e, $p(0) = N_v = \times10^{19} cm^{-3}$, at $x=0$. 
The excellent agreement between simulated $J(V)$ results and experimental J-V characteristics for three different single layer deices based on CuPc and ZnPc  over wide range of temperature is shown  in Fig.5. Following are the observations, (i)  the simulated  result follows $J \propto V^m$ dependence,  (ii)  $m$  is 2 at low electric fields and  increases gradually from  $m=2$  at high electric fields and (iii) the value of $m$ increases with  temperature.  Hence, the experimental data presented in Fig.2 , Fig.3, Fig.4 and Fig.5 alongwith simulation results establish the carrier transport process, which is thermally activated hopping with electric field dependent carrier mobility.

 There are three most important models for the explanation of     particular  temperature dependence of  zero field mobility($\mu(0,T)$) and the field activation energy($\gamma(T)$), appearing in universally observed  PF relation(Eq.4.1) about charge  carrier  mobility. Those are, (a) Gill's phenomenological model\cite{wdg72} (b) Uncorrelated Gaussian disorder model\cite{hb93} and (c) Correlated Gaussian disorder model\cite{svn98}. Gill  attempted to describe\cite{wdg72} the experimental data with the following empirical form of charge carrier  mobility
 
\begin{equation}
\mu(F,T)= \mu_{0} \exp \left(-\frac{\Delta}{k_{B}T}\right) \exp
\left[ \frac{B}{k_{B}}\left(\frac{1}{T}-\frac{1}{T_{0}}\right) \sqrt{F} \right ] 
\end{equation}

\noindent where $\mu_{0}$ is the temperature independent mobility, 
$\Delta$ is the zero-field activation energy, $B$ is a parameter,
and $T_0$ is the characteristic temperature of the material.  In this case, $\mu(0,T)$ and $\gamma(0,T)$ have $1/T$ dependence.  Comparing this with the PF relation(Eq.1), $\mu(0,T)$ and $\gamma(0,T)$ can be expressed as

\begin{equation}
\mu(0,T) =\mu_{0}\exp\left(-\frac{\Delta}{k_{B}T}\right), \gamma(T) =\frac{B}{k_{B}}\left(\frac{1}{T}-\frac{1}{T_{0}}\right)
\end{equation} 

\noindent 
The $\mu(0,T)$ and $\gamma(T)$  at different
temperatures,  determined from theoretical fit to
experimental J-V characteristics(shown in Fig.5) are plotted in Fig.6 according to Eq.8. From $ln[(\mu(0,T))]$ vs $1/T$ plot, the values of temperature independent mobility $\mu_0$  and the activation energy $\Delta$ are calculated. The value of the constant B and the characteristics temperature $T_0$  are calculated from the $\gamma(T)$ vs $1/T$ plot and given in Table-I.

Bassler and co-workers proposed\cite{hb93} the uncorrelated Gaussian disorder model(UGDM) and provided support to PF behavior of mobility using Monte Carlo simulation. The UGDM describes the carrier transport as a biased random walk among the hopping sites  with Gaussian-distributed random site energies. UGDM lead to following form of carrier mobility

\begin{equation}
\mu(F,T)=\mu_{0}\exp \left[-\frac{2}{3}\hat{\sigma}^2 +C\left(\hat{\sigma}^2-\Sigma^2\right)\sqrt{F}\right]
\end{equation}

\noindent where $\hat{\sigma}= \sigma /k_BT$, C $\& \; \sum$ are constants,
$\sigma$ is the Gaussian width of the hopping sites distribution. In this case, $\mu(0,T)$ and $\gamma(0,T)$ have $1/T^2$ dependence, instead of $1/T$ dependence as in case of Gill's empirical relation. Comparing this with the PF relation(Eq.1), $\mu(0,T)$ and $\gamma(0,T)$ can be expressed as

\begin{equation}
\mu(0,T) =\mu_{0}\exp \left(-\frac{2}{3}\hat{\sigma}^2\right),
\gamma(T)=C\left(\hat{\sigma}^2-\Sigma^2\right)
\end{equation}

\noindent Fig.7 shows the temperature dependence of the $\mu(0,T)$  and  $\gamma(T)$ according to  UGDM and the apparent linear dependence of ln$\mu(0)$ on $1/T^2$ and $\gamma(T)$ on $1/T^2$ can be described by  Eq.10.
Values  of temperature independent mobility $\mu_0$, width of the energy spread $\sigma$, and the constants $C$ and $\Sigma$ are determined from the  ln$\mu(0,T)$ vs. $1/T^2$ and $\gamma(T)$ vs. $1/T^2$ plots, and given in Table II. 
Although UGDM explains some features of experimental data and provides support for PF behavior of carrier mobility, several discrepancies emerge with uncorrelated description of Gaussian disorder model, which will be discussed later.  The most important criticism against UGDM is its' inability to reproduce the PF behavior over wider range of electric field. 
Garstein and Conwell\cite{yng95} first showed that a spatially correlated potential is required for the description of PF behavior of mobility for wider range of electric field.  Dunlop and co-workers\cite{dhd96} have shown  that the interaction of charge carriers with permanent dipoles located on either dopant or host molecules give rise to PF behavior of mobility. Essentially, this correlated Gaussian disorder model(CGDM) is based on long-range correlation\cite{svn98} between charge carriers and the molecular electric dipole, resulting  random potential energy landscape with long-range spatial correlations, $<U(0)U(r)> \sim \sigma^2a/r$,  where $a$ is the minimal charge-dipole separation or the lattice constant of the molecular solid and $\sigma$ is the width of the Gaussian distribution and can be given by $\sigma=2.35 e p/\epsilon a^2$, where $p$ is the dipole moment of the molecule.  In this case,  carrier mobility can be given by  

\begin{equation}
\mu(F,T)  =\mu_{0} exp\left[-\left\{\frac{3}{5}\hat{\sigma}\right\}^2 + A \left(\hat{\sigma}^{3/2}-\Gamma \right)\right] \sqrt{\frac{eaF}{\sigma}} 
\end{equation}

\noindent where  $A$ and $\Gamma$ are parameters of the model. $\Gamma$ characterizes the geometrical disorder. In this case,  $\mu(0,T)$ has a $1/T^2$ dependence(same as in UGDM) but $\gamma(T)$ has a $1/T^{3/2}$ dependence.  Comparing this with the PF relation(Eq.1), $\mu(0,T)$ and $\gamma(0,T)$ can be expressed as

\begin{equation}
\mu(0,T)=\mu_{0} exp\left[-\left(\frac{3}{5}\hat{\sigma}\right)^2\right], 
\gamma(T)=C_0 \left( \hat{\sigma}^{3/2}-\Gamma \right)\sqrt{\frac{ea}{\sigma}} 
\end{equation}

\noindent Fig.8 shows the temperature dependence of $\mu(0,T)$ and $\gamma(T)$ according to CGDM, described by Eq.12. Values of the parameters $\mu_0$, $\sigma$, $a$ and $\Gamma$ are determined from the straight line plots of ln$\mu(0,T)$ vs. $1/T^2$(shown in Fig.8) and $\gamma(T)$ vs. $1/T^{3/2}$(shown in Fig.8) and given in Table III.

Though the uncorrelated models(Gill's phenomenological relation and UCGM) can  describe the experimental data upto a some  extent, but have certain limitations which  are discussed below. There are several conceptual problems with Gill's phenomenological description and the  most important are (i) the obtained values of zero field and temperature independent mobility($\mu_0$) are overestimated by several order of magnitudes and don't match with the values obtained by independent measurements, (ii) the empirical Gill's formula lacks theoretical justification and the significance of different parameters($\Delta$, $B$ and $T_0$) are not well understood in the context of the physical properties of the system and (iii) the empirical formula for $\gamma(T)$ predicts the negative field dependence for $T\geq T_0$, which deviates from the universal feature(PF) of carrier transport in these materials.

Main feature of UGDM is the non-Arrehenious behavior of the temperature dependence of mobility, which is the consequence of hopping conduction in GDOS. But, there are several discrepancies in this description and those are  (i) fitting is not as good as in case of CGDM, (ii) the intercept of $\gamma(T)$ vs. 1/T$^2$(shown in Fig.7) gives positive value, but according to UGDM relation Eq.10, it should be negative, which is a fundamental problem with UGDM, (iii) it has been observed that there is no specific trend in the values of $C$, which is scaling parameter in the model and cannot be linked to a physical parameter of the system and (iv) its' inability to reproduce the PF behavior(which is a universal feature of charge carrier transport in these materials.) over wider range of electric field\cite{lbs92}.

 Here, we compare our experimental data  with the uncorrelated(UGDM) and correlated(CGDM) theoretical models proposing the stretched exponential electric field  dependence of $\mu$.  As already mentioned, at every temperature, $\mu(0,T)$ and  $\gamma(T)$ are obtained from the experimental data presented in Fig.5,  using the numerical workout  and plotted in Fig.8  which shows the temperature dependence of $\mu(0,T)$ and $\gamma(T)$ according to CGDM, described by Eq.12.   Following observations have been made regarding the CGDM description of the experimental data, (i) linear dependence of ln$\mu(0,T)$ on $1/T^2$ and $\gamma(T)$ on $1/T^{3/2}$ is evident in Fig.8  and straight line fit is excellent and better than that in UGDM,  (ii) for all samples the intercept of $\gamma(T)$ vs. $1/T^{3/2}$ gives negative values, according to CGDM(Eq.12), (iii) the experimental value of $\Gamma$ are 1.2 in CuPc and 1.4 in ZnPc, which are very close to the predicted value 1.97\cite{snn98}, (iv) The width of the Gaussian distribution $\sigma$ is 100meV in CuPc and  120meV in  ZnPc and similar values for $\sigma$ are commonly observed in other molecular solids, (v) The lattice constant($a$) has been found to be 19$\AA$ in CuPc and 17$\AA$ in ZnPc, are in excellent agreement with the reported\cite{rr00} experimental values determined by X-Ray diffraction.

The UCDM and CGDM share the common feature  regarding the temperature dependence of  $\mu(0,T)$ due to similar distribution of hopping sites energies, but the temperature dependence of field activation   parameter $\gamma$ decides critically the importance of spatial correlation and decides the mechanism of carrier transport in organic molecular solids. Though CGDM explains the experimental data successfully, there are some limitations with this model.  The positional disorder is not included in the present form of CGDM, but large values of disorder parameters  $\sigma$ and $\Gamma$  in NPPDA\cite{svn98}, Alq3\cite{ggm01} and  MePc  signify the importance of the positional disorder. In CGDM, it has been shown that long-range interaction between charge carriers and permanent dipole moments of doped molecules in polymers and  host molecules leads to spatial correlation. However, it has been pointed out by Yu {\sl et. al.}\cite{zgy00} that the mechanism responsible for PF behavior in different conjugated polymers and molecules cannot be due to only charge-dipole interaction, because  PF behavior has been universally observed in several doped and undoped polymers and molecules with or without permanent dipole moment.  Hence, in addition to charge-dipole interaction there may be another mechanism responsible for spatial correlation, which is a fundamental requirement for PF behavior. Yu {\sl et. al.}\cite{zgy00} have  shown using first principle quantum chemical calculation that the thermal fluctuations in the molecular geometry can lead to spatial correlation. It has been shown that the primary restoring force for the thermal fluctuation is steric and intermolecular, which lead to spatially correlated fluctuation in the energies of the localized states. Further experimental investigations are in progress to identify the exact origin of spatial correlation in molecules with and without dipole moments.

In summary, we have studied the mechanism of carrier transport in OMS, which belongs to class of disordered molecular solids, by means of simple  experiments. We have (i) shown that the carriers conducts through hopping; (ii) given evidences for the universal electric field dependence($ln(\mu) \propto \sqrt{E}$) of carrier mobility and (iii) presented a comparison of the transport properties with uncorrelated and correlated disorder model. We have shown  that $E$ and $T$ dependence of experimental data can be fitted and described within the correlated Gaussian disorder model, signifying the important role of both energetic and spatial disorder on the carrier transport in these materials. The origin of correlation has been discussed. We hope present analysis would lead to further experimental and theoretical investigations focusing the exact origin and role of correlation on transport properties of disordered molecular solids.

\newpage

\noindent {\large \bf Figure Captions}

\begin{description}

\vspace{0.3in}

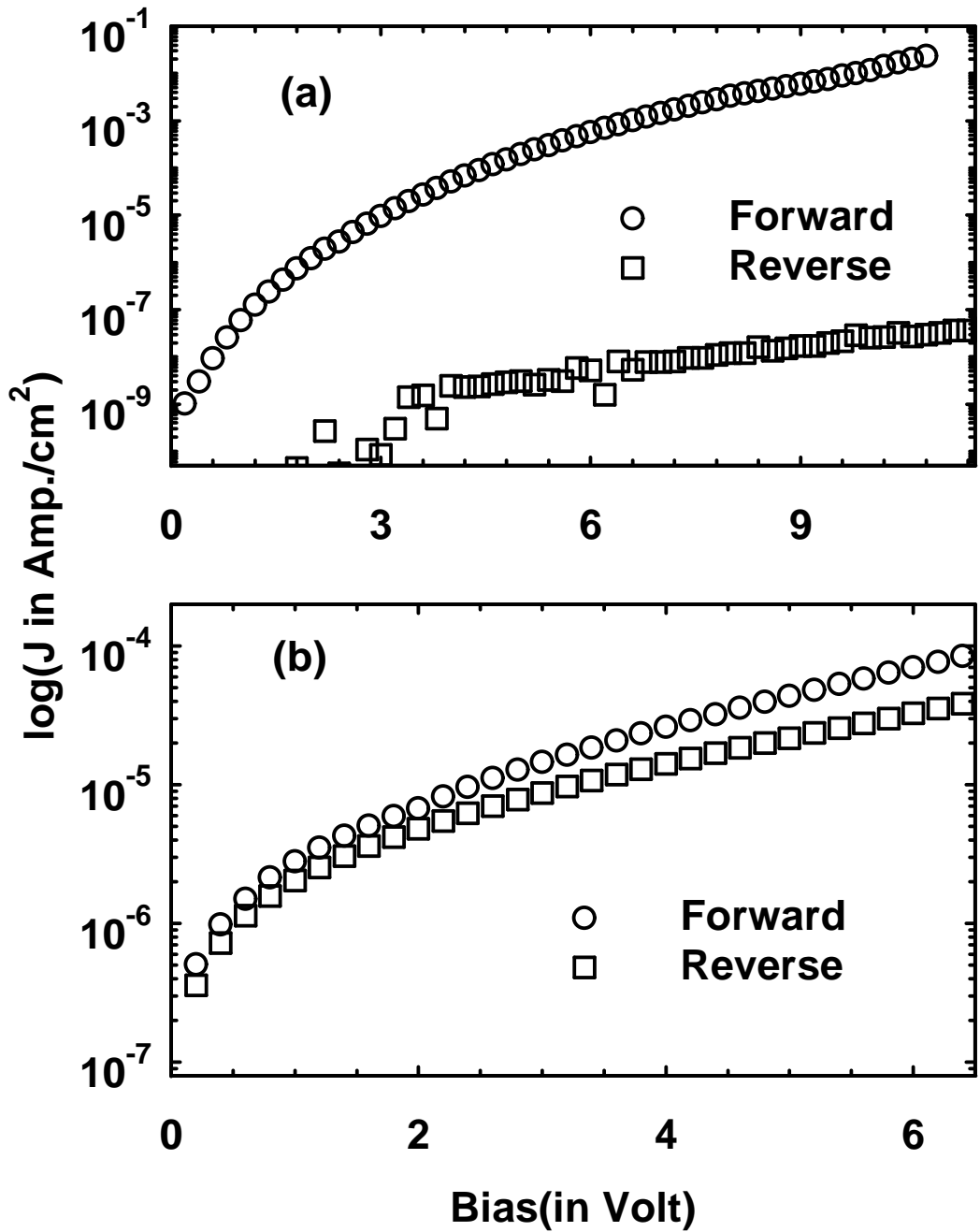
\item[Fig.1.] Room temperature J-V characteristics for single layer hole only CuPc device with thickness of 200nm.  Empty circles represent the data when  ITO is positively biased (Forward biased)and empty squares represent the data when ITO is negatively biased(reverse biased). The high rectification in (a) ITO/CuPc/Al and comparable currents in (b) ITO/CuPc/Cu ensures the condition to achieve the bulk limited current is when ITO is positively biased.

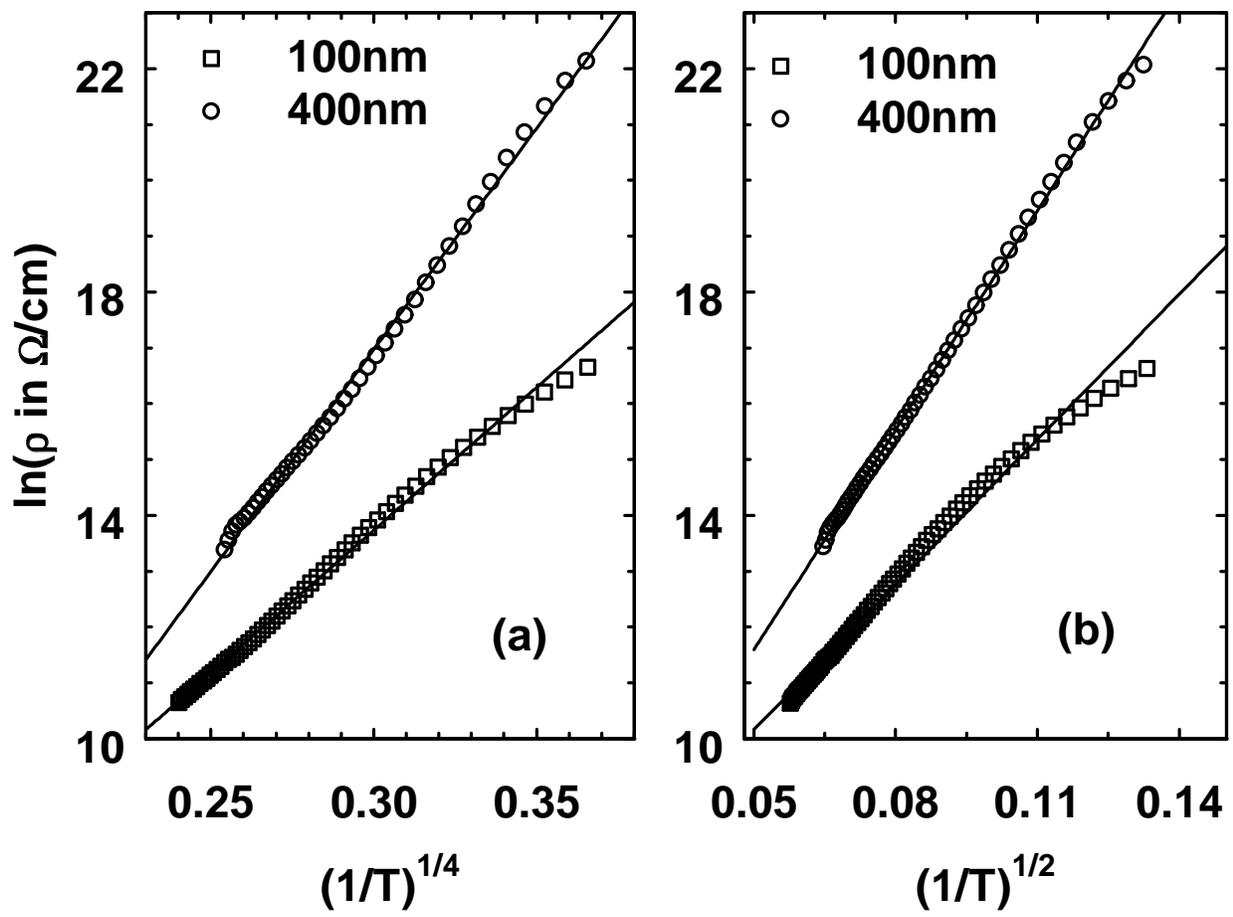
\item[Fig.2]  Temperature dependent resistivity of CuPc observed in ITO/CuPc/Al structure with 100nm and 400nm of CuPc layers.  Empty symbols are experimental data and solid lines are the straight line fit in the log($\rho$) vs. $1/T^{1/4}$ in (a) and vs $1/T^{1/2}$  in (b).

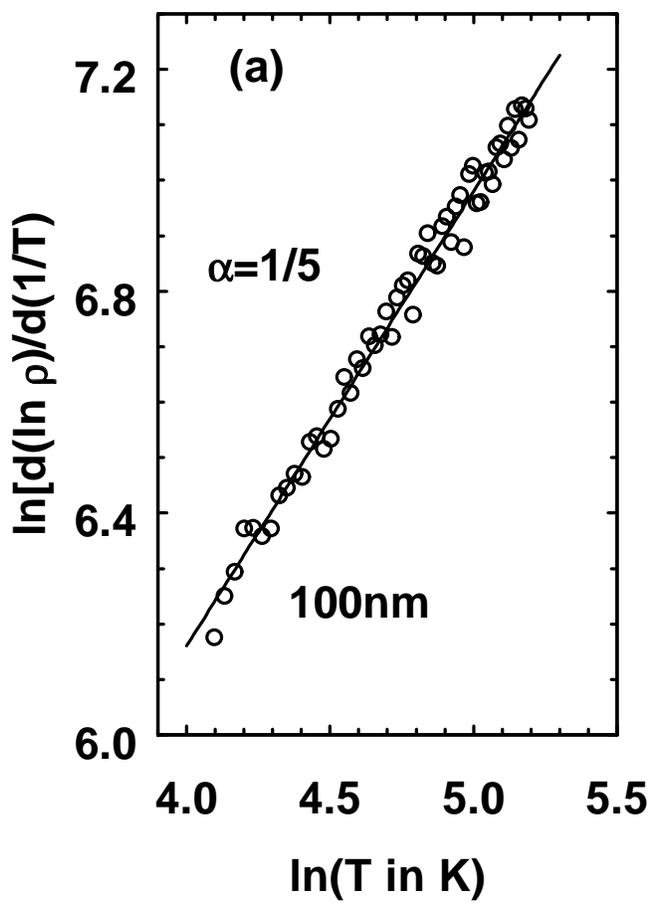
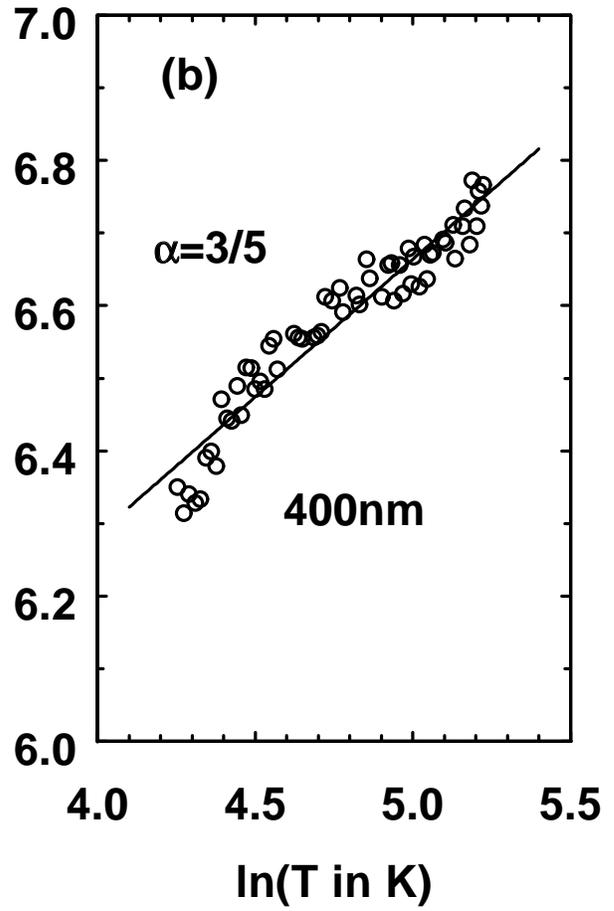
\item[Fig.3.] Temperature dependent resistivity of CuPc in ITO/CuPc/Al structure plotted according to Eqn.4 for (a)100nm and (b)400nm CuPc layers. From the straight line fit(solid line) to the experimental data(symbols)), the values of $\alpha = 1/5$ for 100nm and $\alpha = 3/5$ for 400nm structures were found out.

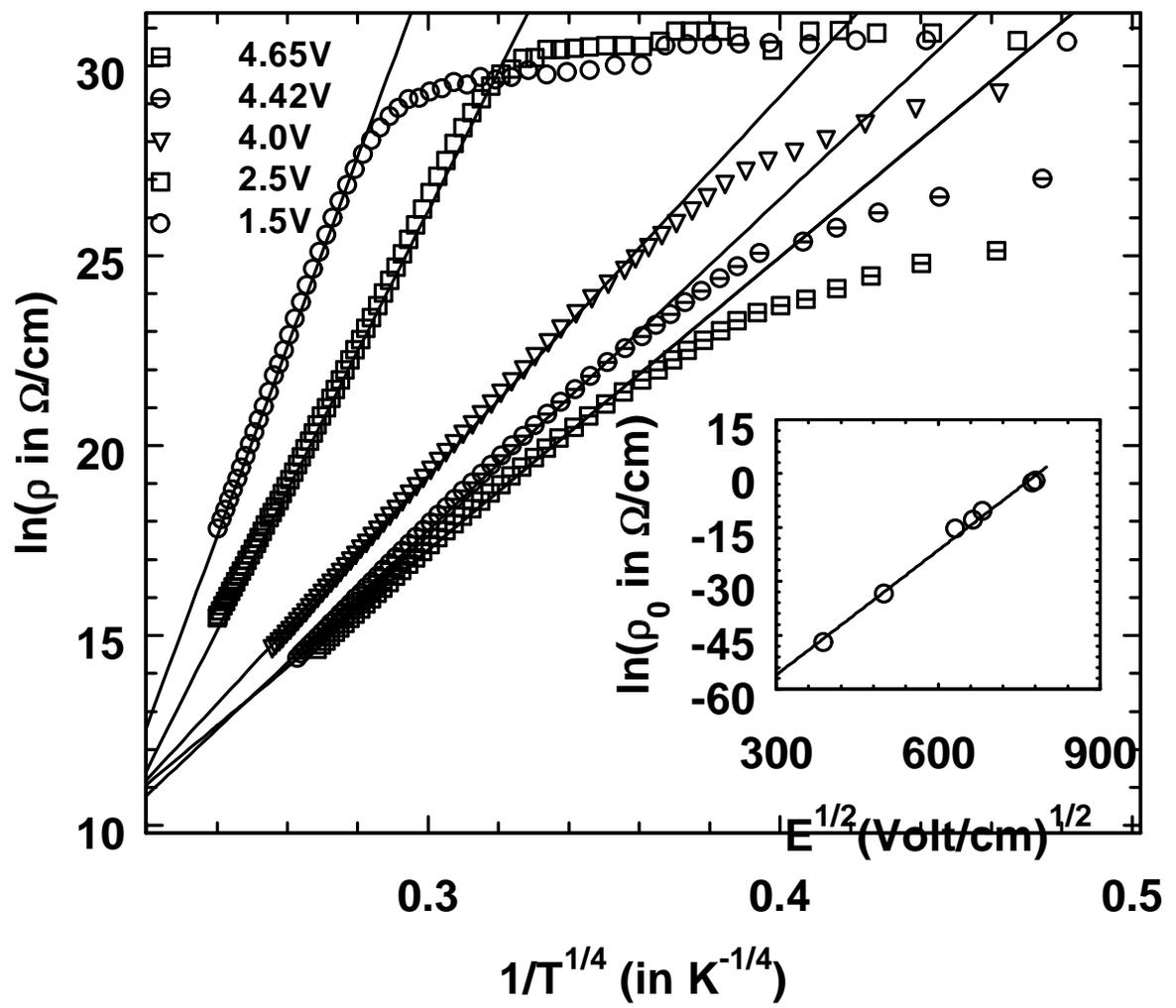
\item[Fig.4.]  Temperature dependent conductivity of CuPc observed in ITO/CuPc(100nm)/Al structure.  Empty circles are experimental data and solid lines are the straight line fit in the log($\rho$) vs. $1/T^{1/4}$. Inset shows the electric field dependence of  temperature independent resistivity($\rho_0$) at different electric fields and the linear fit(solid line) shows the PF dependence of the carrier mobility.

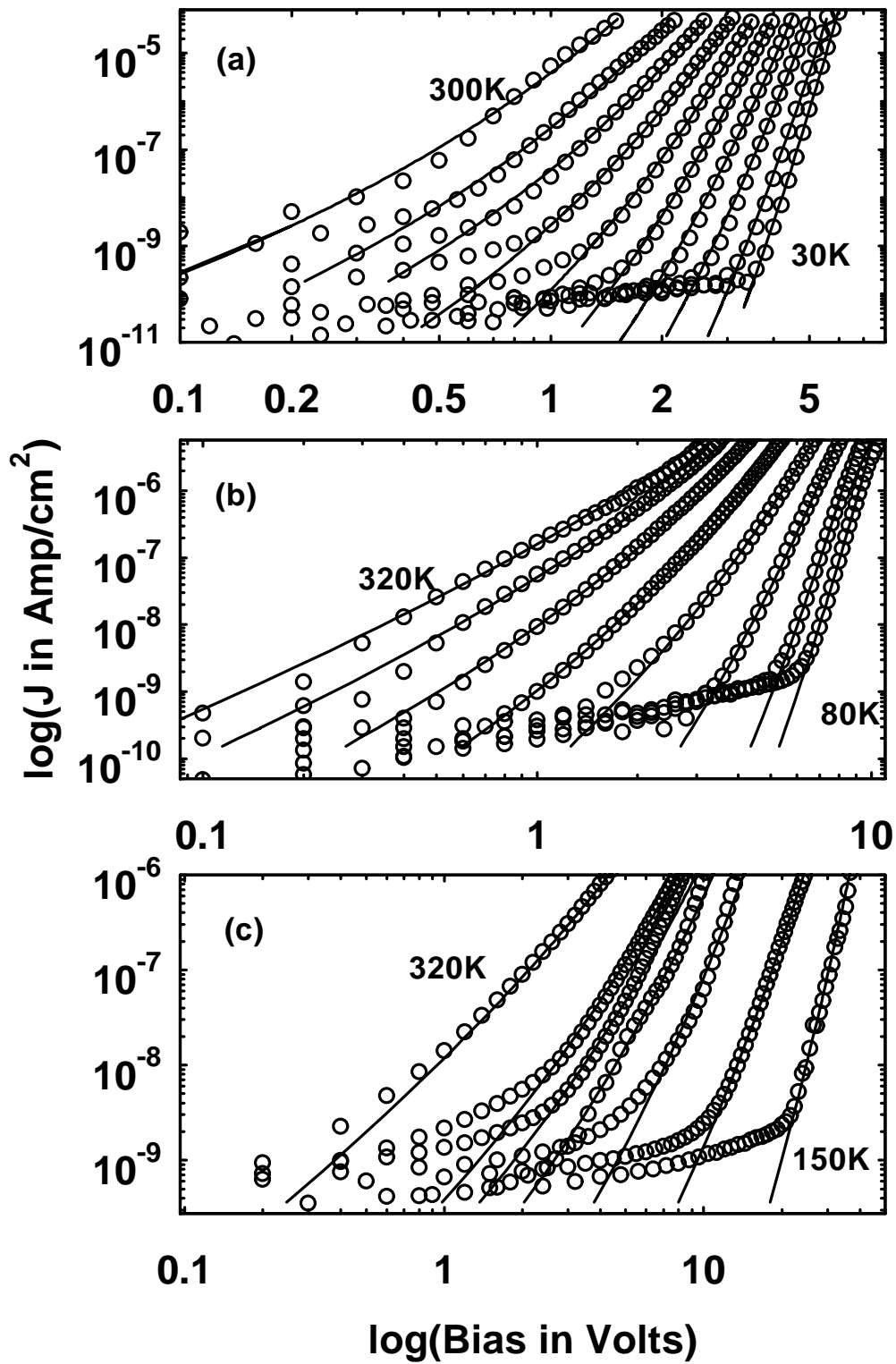
\item[Fig.5.] J-V characteristics of  bulk limited current in ITO/MePc/Al with ITO anode  and Al cathode contacts  at different temperatures starting at 320K and then at the interval of 30K. Experimental data are shown by the empty symbols and solid lines represent the  simulated data, for (a) 100nm CuPc layer(device A), (b) 200nm CuPc layer(device B), and (c) 200nm ZnPc layer(device C).

\item[Fig.6.] (a) Temperature dependence of zero field mobility $\mu(0,T)$ and (b) the field activation $\gamma(T)$ obtained from the theoretical fit to the experimental data shown in Fig.5, are plotted according to the Gill's phenomenological observation(Eq.8) for devices A, B, and C. The values of the parameters $\mu_0$, $\Delta$, $B$, and $T_0$  are given in Table-I.

\item[Fig.7.] Temperature dependence of zero field mobility $\mu(0,T)$ and the field activation $\gamma(T)$ obtained from the theoretical fit to the experimental data shown in Fig.5, are plotted according to the UGDM(Eq.10) for devices A, B, and C. The values of the parameters $\mu_0$, $\sigma$, $C$ and $\Sigma$ are given  in Table-II.

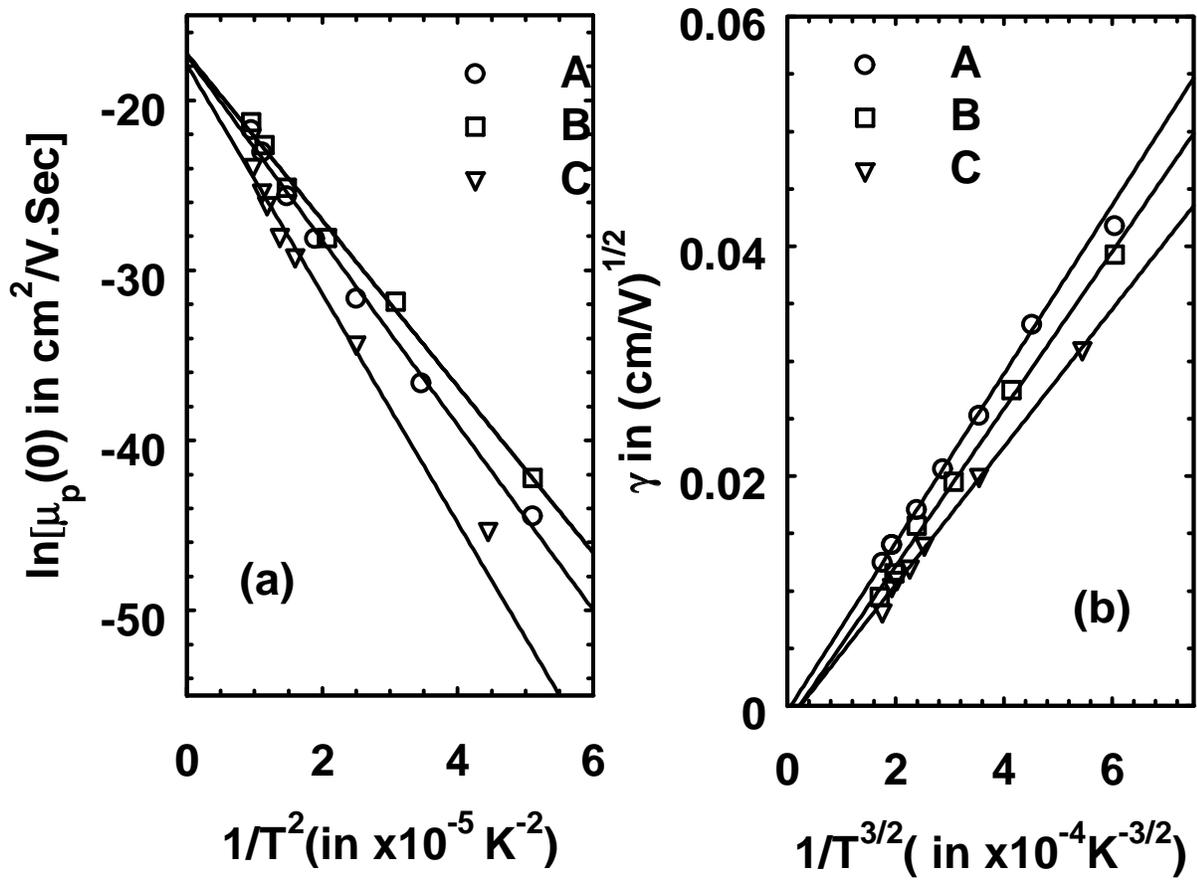
\item[Fig.8.] Temperature dependence of zero field mobility $\mu(0,T)$ and the field activation $\gamma(T)$ obtained from the theoretical fit to the experimental data shown in Fig.5, are plotted according to the CGDM(Eq.12) for devices A, B, and C. The values of the parameters $\mu_0$, $(\sigma)$, and $a$ are given  in Table-III. 

\end{description}

\vspace{0.5in}

\noindent {\large \bf Table Captions}

\vspace{0.3in}

\begin{description}

\item[Table.I]  Values of $\mu_0$,  $\Delta$, B and $T_0$ obtained using Gill's phenomenological relation for three different samples.

\item[Table.II] Values of $\mu_0$,  $\sigma$, C and $\Sigma$ obtained using UGDM for three different samples.

\item[Table.III]  Values of $\mu_0$, $\sigma$ and $a$ using CGDM for three different samples.

\end{description}

\newpage

\begin{center}
{\bf Table-I}
\end{center}

\begin{center}
	\begin{tabular}{|ccccc|}\hline 
 ~~Device~~ & $\mu_0$ & $\Delta$ & B & $T_0$ \\ 	
& ~~in $\frac{cm^2}{Vs}$~~ & ~~in eV~~ & ~~in eV(cm/V)$^{\frac{1}{2}}$~~ & ~~in K~~ \\  \hline
A	 & $2.76 \times 10^{-3}$ & 0.44 & $6.38\times 10^{-4}$ & 660 \\	
B	 & $3.78 \times 10^{-3}$ & 0.43 & $5.94\times 10^{-4}$ & 615 \\	
C	 & $6.1 \times 10^{-4}$ & 0.46 & $5.28\times 10^{-4}$ & 610 \\	\hline
	\end{tabular}
\end{center}

\vspace{1in}

\begin{center}
{\bf Table-II}
\end{center}

\begin{center}
\begin{tabular}{|ccccc|}\hline 
 ~~Device~~ & $\mu_0$ & $\sigma$ & C & $\Sigma$\\
 & ~~(in $cm^2/V.sec)$~~ & ~~(in meV)~~ & &  \\  \hline
A	 & $2.8\times 10^{-8}$ & 100 & ~~$6.4\times 10^{-5}$~~ & ~~-3.1~~\\	
B	 & $3.0\times 10^{-8}$ & 95 & $ 6.9\times 10^{-5}$ & -3.18\\	
C	 & $1.7\times 10^{-8}$ & 110 & $ 4.5\times 10^{-5}$ & -4.24\\  \hline
	\end{tabular}
\end{center}

\vspace{1in}

\begin{center}
{\bf Table-III}
\end{center}

\begin{center}
	\begin{tabular}{|cccc|} \hline 
 ~~Device~~ & $\mu_0$ & $\sigma$ & a \\ 
	 & ~~(in $cm^2/V.sec$)~~& ~~(in meV)~~ & ~~(in $\AA$)~~ \\	 \hline
A	 & $2.8\times 10^{-8}$ & 106 & 20 \\	
B	 & $3.0\times 10^{-8}$ & 100 & 19\\			
C	 & $1.7\times 10^{-8}$ & 118 & 17 \\		\hline		
	\end{tabular}
\end{center}

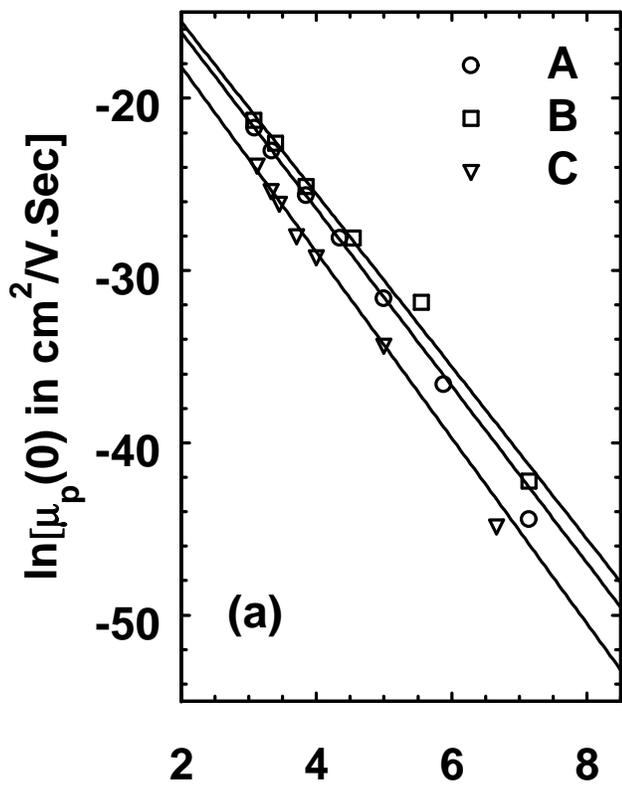
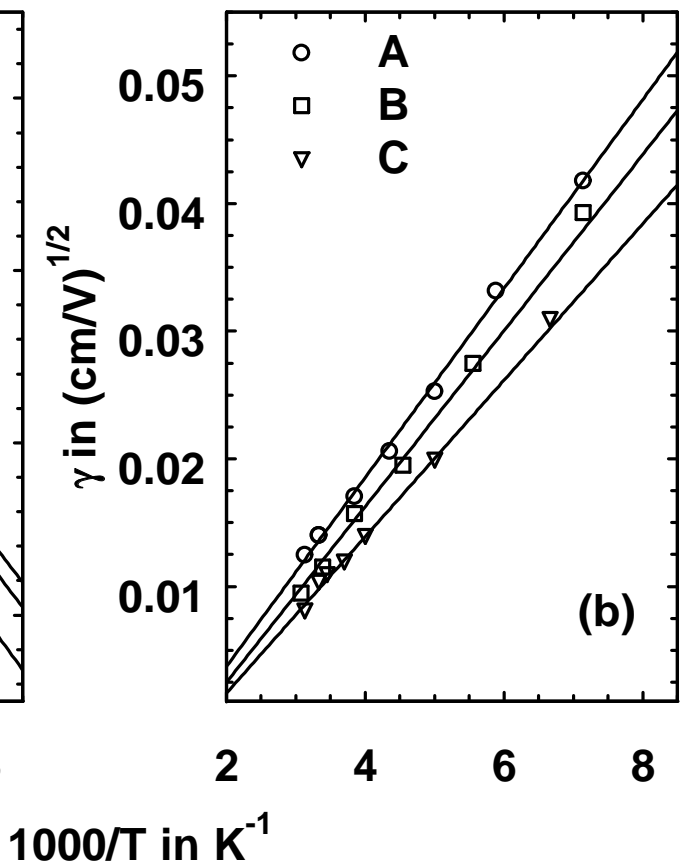
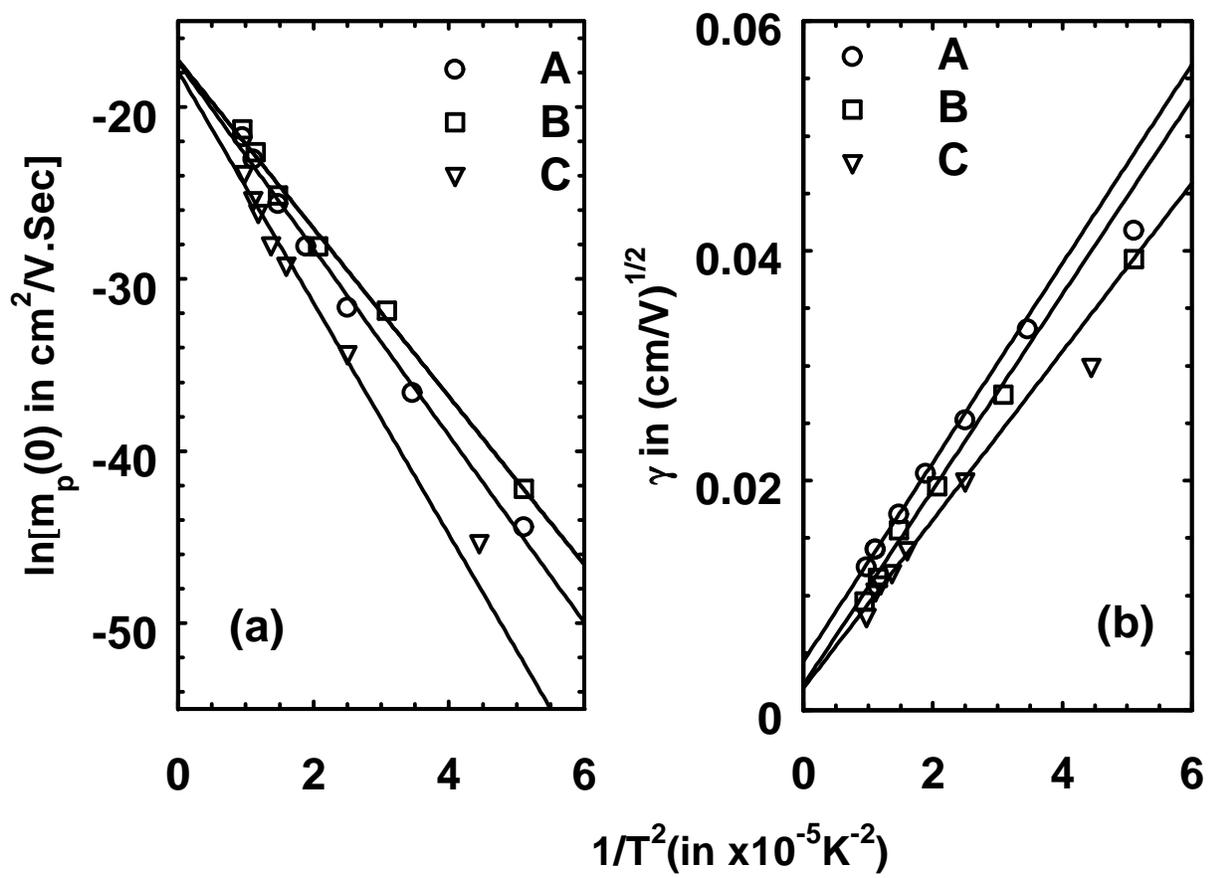
\end{document}